\documentclass[a4paper]{report}
\usepackage[utf8]{inputenc}
\usepackage[T1]{fontenc}
\usepackage{RJournal}
\usepackage{amsmath,amssymb,array}
\usepackage{booktabs}

\begin{document}

\sectionhead{Contributed research article}
\volume{XX}
\volnumber{YY}
\year{20ZZ}
\month{AAAA}

\begin{article}
\title{\pkg{Counterfactual}: An R Package for Counterfactual Analysis}
\author{by Mingli Chen, Victor Chernozhukov, Iván Fernández-Val and Blaise Melly}

\maketitle

\abstract{
The \pkg{Counterfactual} package implements the estimation and inference methods of \citep{ChernozhukovFernandez-ValMelly2013} for counterfactual analysis. The counterfactual  distributions considered are the result of changing either the marginal distribution of covariates related to the outcome variable of interest, or the conditional distribution of the outcome given the covariates. They can be applied to estimate quantile treatment effects and wage decompositions. This paper serves as an introduction to the package and displays basic functionality of the commands contained within. 
}

\section{Introduction \label{sec:Introduction}}

Using econometric terminology, we can often think of a counterfactual distribution as the result  of a change in either the distribution of a set of covariates $X$ that determine the outcome variable of interest $Y$, or the relationship of the covariates with the outcomes, that is, a change in the conditional distribution of $Y$ given $X$. Counterfactual analysis consists of evaluating the effects of such changes.  The \pkg{Counterfactual} package implements the methods of \citep{ChernozhukovFernandez-ValMelly2013} for counterfactual analysis. It contains commands to estimate and make  inference on quantile effects constructed from counterfactual distributions. The counterfactual distributions are estimated using regression methods such as classical, duration, quantile and distribution regressions.  The inference on the quantile effect function can be pointwise at a specific quantile index or uniform  over a range of specified quantile indexes. 

\medskip

We start by giving a simple example of counterfatual analysis. Suppose we would like to analyze the wage differences between men and women. Let $0$ denote the population of men and let $1$ denote the population of women. The variable $Y_{j}$ denotes wages and $X_{j}$ denotes job market-relevant characteristics that affect wages for populations $j=0$ and $j=1$. The conditional distribution functions $F_{Y_0|X_0}(y|x)$ and $F_{Y_1|X_1}(y|x)$ describe the stochastic assignment of wages to workers with characteristics $x$, for men and women, respectively. Let $F_{Y\left\langle 0|0\right\rangle }$ and $F_{Y\left\langle 1|1\right\rangle }$ represent the observed distribution function of wages for men and women, and let $F_{Y\left\langle 0|1\right\rangle }$ represent the distribution function of wages that would have prevailed for women had they faced the men's wage schedule $F_{Y_0|X_0}$:
\[
F_{Y\left\langle 0|1\right\rangle }(y):=\int_{\mathcal{X}_{1}}F_{Y_{0}|X_{0}}(y|x)dF_{X_{1}}(x).\label{eq:DefCounter}
\]
The latter distribution is called counterfactual, since it does not arise as a distribution from any observable population. Rather, this distribution is constructed by integrating the conditional distribution of wages for men with respect to the distribution of characteristics for women. This quantity is well defined if $\mathcal{X}_0$, the support of men's characteristics, includes $\mathcal{X}_1$, the support of women's characteristics, namely $\mathcal{X}_1\subset \mathcal{X}_0$.

\medskip

Let $F^{\leftarrow}$ denote the quantile or left-inverse function of the distribution function $F$.  The difference in the observed wage quantile function between men and women can be decomposed in the spirit of  \citep{Oaxaca1973} and \citep{Blinder1973} as
\begin{equation}
F^{\leftarrow}_{Y\left\langle 1|1\right\rangle }-F^{\leftarrow}_{Y\left\langle 0|0\right\rangle}=[F^{\leftarrow}_{Y\left\langle 1|1\right\rangle}-F^{\leftarrow}_{Y\left\langle 0|1\right\rangle}]+[F^{\leftarrow}_{Y\left\langle 0|1\right\rangle }-F^{\leftarrow}_{Y\left\langle 0|0\right\rangle }], \label{eq:decomp}
\end{equation}
where the first term in brackets is due to differences in the wage structure and the second term is a composition effect due to differences in characteristics.  These counterfactual effects are well defined econometric parameters and are widely used in empirical analysis, for example, the first term of the decomposition is a measure of gender wage discrimination. In Section \ref{sec:examples} we consider an empirical example where $0$ denotes the population of nonunion workers and $1$ denotes the population of union workers. In this case the the wage structure effect corresponds to the treatment effect of union or union premium. It is important to note that these effects do not necessarily have a causal interpretation without additional conditions that are spelled out in \citep{ChernozhukovFernandez-ValMelly2013}. 



\section{The Counterfactual package}

\subsection{Getting Started}
To get started using the package \pkg{Counterfactual} for the first time, issue the command
\begin{Schunk}
\begin{Sinput}
> install.packages("Counterfactual")
\end{Sinput}
\end{Schunk}
into your R browser to install the package in your computer. Once the package has been installed,
you can use the package \pkg{Counterfactual} during any R session by simply issuing the command
\begin{Schunk}
\begin{Sinput}
> library(Counterfactual)
\end{Sinput}
\end{Schunk}
Now you are ready to use the function \emph{counterfactual} and data sets contained in \pkg{Counterfactual}. For general questions about the package you may type
\begin{Schunk}
\begin{Sinput}
> help(package = "Counterfactual")
\end{Sinput}
\end{Schunk}
to view the package help file, or for more questions about a specific function you can type \samp{help(function-name)}. For example, try:
\begin{Schunk}
\begin{Sinput}
> help(counterfactual)
\end{Sinput}
\end{Schunk}
or simply type
\begin{Schunk}
\begin{Sinput}
> ?counterfactual
\end{Sinput}
\end{Schunk}

The command \emph{counterfactual}  has the general syntax: 

\begin{Schunk}
\begin{Sinput}
> counterfactual(formula, data, weights, na.action = na.exclude, 
+                group, treatment = FALSE, decomposition = FALSE,
+                transformation = FALSE, counterfactual_var, 
+                quantiles, method = "qr", 
+                trimming = 0.005, nreg = 100, scale_variable, 
+                counterfactual_scale_variable, 
+                censoring = 0, right = FALSE, nsteps = 3, 
+                firstc = 0.1, secondc = 0.05, noboot = FALSE, 
+                weightedboot = FALSE, seed = 8, robust = FALSE, 
+                reps = 100, alpha = 0.05, first = 0.1, 
+                last = 0.9, cons_test = 0, printdeco = TRUE, 
+                sepcore = FALSE, ncore=1)
\end{Sinput}
\end{Schunk}

To describe the different options of the command we need to provide some background on methods for counterfactual analysis. 

\subsection{Setting for Counterfactual Analysis}
Consider a general setting with two populations labeled
by $k\in \mathcal{K} = \{0,1\}$. For each population $k$ there is the $%
d_{x}$-vector $X_{k}$ of covariates and the scalar outcome $Y_{k}$.
The covariate vector is observable in all populations, but the outcome is
only observable in populations $j\in \mathcal{J}\subseteq \mathcal{K}$.
Let $F_{X_{k}}$ denote the covariate distribution in population $k\in \mathcal{K},$ and  $%
F_{Y_{j}|X_{j}}$ and $Q_{Y_{j}|X_{j}}$ denote the conditional distribution and quantile functions in population $j\in \mathcal{J}$.
 We denote the support of $X_{k}$ by $%
\mathcal{X}_{k}\subseteq \mathbb{R}^{d_{x}}$, and the region of interest for $%
Y_{j}$ by $\mathcal{Y}_{j}\subseteq \mathbb{R}$. The refer to $j$ as the reference population(s) and to $k$ as the counterfactual population(s).

\medskip

The reference and counterfactual populations in the wage examples correspond to different groups such as men and women or nonunion and union workers. We can also generate counterfactual populations by artificially transforming a reference population. Formally, we can think of $X_k$ as being created through a known transformation of $X_j$:
\begin{equation}
X_{k}=g_{k}(X_{j}),\quad\mathrm{where}\quad g_{k}:\mathcal{X}_{j}\rightarrow\mathcal{X}_{k}.\label{eq:transform}
\end{equation}
This case covers adding one unit to the first covariate, $X_{1,k}=X_{1,j}+1$, holding the rest of the covariates constant. The resulting quantile effect becomes the \textit{unconditional} quantile regression, which measures the effect of a unit change in a given covariate component on the unconditional quantiles of $Y$. For example, this type of counterfactual is useful for estimating the treatment effect of smoking during pregnancy on infant birth weights. Another possible transformation is a mean preserving redistribution of the first covariate implemented as $X_{1,k} = (1-\alpha) E[X_{1,j}] + \alpha X_{1,j}$. These and more general types of transformation defined in (\ref{eq:transform}) are useful for estimating the effect of a change in taxation on the marginal distribution of food expenditure or the effect of cleaning up a local hazardous waste site on the marginal distribution of housing prices (\citep{Stock1991}).  We give an example of this type of transformation in Section \ref{Sec:example2}. 

\medskip

The reference and counterfactual populations can be specified to \emph{counterfactual} in two ways that accommodate the previous two cases:
\begin{enumerate}

\item If the option \emph{group} has been specified, then $j$ is the population defined by \emph{group} and $k$ is the population defined by \emph{group=1}. This means that both $X$ and $Y$ are observed in \emph{group=0}, but only $X$ needs to be observed in \emph{group=1}. When both $X$ and $Y$ are observed in \emph{group=1}, the option \emph{treatment=TRUE} specifies that the structure or treatment effect should be computed, whereas the default option \emph{treatment=FALSE} specifies that the composition effect should be computed; see the definition of the structure and composition effects in the decomposition (\ref{eq:decomp}). If in addition to  \emph{treatment=TRUE} the option \emph{decomposition=TRUE} is selected, then the entire decomposition (\ref{eq:decomp}) is reported including the composition, structure and total effects. Note that we can reverse the roles of the populations defined by an indicator variable \emph{vargroup} by setting either \emph{group=vargroup} or \emph{group=1-vargroup}. 

\item Alternatively, the option \emph{counterfactual\_var} can be used to specify the covariates in the counterfactual population. In this case, the names on the right handside of \emph{formula} contain the variables in $X_j$ and \emph{counterfactual\_var} contains the variables in $X_k$. The option \emph{transformation=TRUE} should be used when $X_k$ is generated as a transformation of $X_j$, e.g., equation (\ref{eq:transform}). The list passed to \emph{counterfactual\_var} must contain exactly the same number of variables as the list of independent variables in \emph{formula} and  the order of the variables in the list matters.
\end{enumerate}

Counterfactual distribution and quantile functions are formed by combining the conditional distribution in the population $j$
with the covariate distribution in the population $k$, namely: 
\begin{eqnarray*}
& F_{Y{\langle j|k\rangle }}(y):=\int_{\mathcal{X}%
_{k}}F_{Y_{j}|X_{j}}(y|x)dF_{X_{k}}(x),\ \ y\in \mathcal{Y}_{j},
\label{define: counter} \\
& Q_{Y{\langle j|k\rangle }}(\tau ):=F_{Y{\langle j|k\rangle }}^{\leftarrow
}(\tau ),\ \ \tau \in (0,1),
\end{eqnarray*}%
where $(j,k) \in \mathcal{J}\mathcal{K}$, and $F_{Y{\langle j|k\rangle }}^{\leftarrow }(\tau) = \inf \{y \in \mathcal{Y}_{j}: F_{Y{\langle j|k\rangle }}(y) \geq \tau \} $ is the left-inverse
function of $F_{Y{\langle j|k\rangle }}$.  The main interest lies in the quantile effect (QE) function, defined as the difference of two counterfactual quantile  
functions over a set of quantile indexes $\mathcal{T} \subset (0,1)$: 
\[
\Delta(\tau) =  Q_{Y{\langle j|k\rangle }}(\tau ) - Q_{Y{\langle j|j \rangle }}(\tau ), \ \tau \in \mathcal{T},  
\]
where $j \in \mathcal{J}$ and $k \in \mathcal{K}$. In the example of Section \ref{sec:Introduction}, we obtain the composition effect with $j=0$ and $k=1$. When $Y_k$ is observed, then we can construct the structure effect or treatment effect on the treated
\[
\Delta(\tau) =  Q_{Y{\langle k|k \rangle }}(\tau ) - Q_{Y{\langle j|k \rangle }}(\tau ), \ \tau \in \mathcal{T},  
\]
by specifying the option \emph{group} and setting \emph{treatment=TRUE}. In the example of Section \ref{sec:Introduction}, we obtain the wage structure effect with $j=0$ and $k=1$, i.e. setting \emph{group=1} and \emph{treatment=TRUE}. If in addition we select the option \emph{decomposition=TRUE}, then we obtain the entire decomposition (\ref{eq:decomp}) including the composition, structure and total effects. The total effect is
\[
\Delta(\tau) =  Q_{Y{\langle k|k \rangle }}(\tau ) - Q_{Y{\langle j|j \rangle }}(\tau ), \ \tau \in \mathcal{T}.  
\]
The set $\mathcal{T}$ is specified with the option \emph{quantiles}, which enumerates the quantile indexes of interested and should be a vector containing numbers between 0 and 1.  

To estimate the QE function we need to model and estimate the conditional distribution $F_{Y_{j}|X_{j}}$ and covariate distribution $F_{X_k}$. 
We estimate  the covariate distribution using the empirical distribution, and consider several regression based methods for the conditional distribution including classical, quantile, duration, and distribution regression.  Given the estimators of the conditional and covariate distributions $\hat{F}_{Y_{j}\vert X_{j}}$ and $\hat{F}_{X_{k}}$, the estimator of each counterfactual distribution is obtained 
by the
plug-in rule, namely $$\hat{F}_{Y\left\langle j|k\right\rangle}(y)=\int_{\mathcal{X}_{k}}\hat{F}_{Y_{j}\vert X_{j}}(y\vert x)d\hat{F}_{X_{k}}(x),
 y\in\mathcal{Y}_{j}.$$ 
Then, the estimator of the QE function is also obtained by the plug-in rule as
$$\hat \Delta(\tau) =  \hat{F}_{Y\left\langle j|k\right\rangle}^{\leftarrow}(\tau) - \hat{F}_{Y\left\langle j|j\right\rangle}^{\leftarrow}(\tau), \ \ \tau \in \mathcal{T},$$
or
$$\hat \Delta(\tau) =  \hat{F}_{Y\left\langle k|k\right\rangle}^{\leftarrow}(\tau) - \hat{F}_{Y\left\langle j|k\right\rangle}^{\leftarrow}(\tau), \ \ \tau \in \mathcal{T},$$
if we define the counterfactual population with \emph{group} and set \emph{treatment=TRUE}. If in addition to \emph{treatment=TRUE}, we select \emph{decomposition=TRUE}, then the plug-in  estimator of the total effect is
\[
\hat \Delta(\tau) =  \hat{F}_{Y\left\langle k|k\right\rangle}^{\leftarrow}(\tau) - \hat{F}_{Y\left\langle j|j\right\rangle}^{\leftarrow}(\tau), \ \ \tau \in \mathcal{T}.  
\]

\subsubsection{Estimation of Conditional Distribution}
In this section we assume that we have samples $\{(Y_{ji}, X_{ji}): i = 1, \ldots, n_j \}$ composed of independent and identically distributed copies of $(Y_j,X_j)$ for all populations $j \in \mathcal{J}$.
The conditional distribution $F_{Y_j|X_j}$ can be modeled and estimated directly, or throught the conditional quantile function, $Q_{Y_j|X_j}$, using the relation
\begin{equation}
F_{Y_{j}|X_{j}}(y|x)\equiv\int_{(0,1)}1\{Q_{Y_{j}|X_{j}}(u|x)\leq y\}du.\label{eq:QEDEtran}
\end{equation}

The option \emph{formula} specifies the outcome $Y$ as the left hand side variable and the covariates $X$ as the right hand side variable(s). The option \emph{method} allows to select the method to estimate the conditional distribution. The following methods are implemented:

\begin{enumerate}

\item \emph{method = "qr"}, which is the default, implements the quantile regresion estimator of the conditional distribution
\begin{equation}
\hat F_{Y_{j}|X_{j}}(y|x) = \varepsilon + \int_{(\varepsilon,1-\varepsilon)}1\{x'\hat \beta_j(u) \leq y\}du,\label{qr:dist}
\end{equation}
where $\varepsilon$ is a small constant that avoids estimation of tail quantiles, and $\hat \beta(u)$ is the \citep{KoenkerBassettJr1978} quantile regression estimator
$$
\hat \beta_j(u) = \arg \min_{b \in \mathbb{R}^{d_x}} \sum_{i=1}^{n_j} [u - 1\{Y_{ji} \leq X_{ji}'b\}] [Y_{ji} - X_{ji}'b].
$$
The quantile regression estimator calls the  R package \CRANpkg{quantreg} \citep{quantreg-package}. 
The option \emph{trimming} specifies the value of the trimming parameter $\varepsilon$, with default value $\varepsilon = 0.005$.  The option \emph{nreg} sets the number of quantile regressions used to approximate the integral in (\ref{qr:dist}), with a default value of $100$ such that $(\varepsilon,1-\varepsilon)$ is approximated by the grid $\{\varepsilon, \varepsilon + (1-2\varepsilon)/99, \varepsilon + 2 (1-2\varepsilon)/99,  \ldots, 1 - \varepsilon \}$. 
This method should be used only with continuous dependent variables.

\item \emph{method = "loc"}  implements the estimator of the conditional distribution
\begin{equation}
\hat F_{Y_{j}|X_{j}}(y|x) = \frac{1}{n_j} \sum_{i=1}^{n_j} 1 \{Y_{ji} - X_{ji}'\hat \beta_j \leq y - x'\hat \beta_j \},\label{loc:dist}
\end{equation}
where $\hat \beta_j$ is the least square estimator
\begin{equation}
\hat \beta_j = \arg \min_{b \in \mathbb{R}^{d_x}} \sum_{i=1}^{n_j}  (Y_{ji} - X_{ji}'b)^2.\label{eq:ols}
\end{equation}
The estimator (\ref{loc:dist}) is based on a  restrictive location shift model that imposes that the covariates $X$ only affect the location of the outcome $Y$.

\item \emph{method = "locsca"} implements the estimator of the conditional distribution
\begin{equation}
\hat F_{Y_{j}|X_{j}}(y|x) = \frac{1}{n_j} \sum_{i=1}^{n_j} 1 \left\{\frac{Y_{ji} - X_{ji}'\hat \beta_j}{\exp (X_{2ji}'\hat \gamma_j/2)} \leq \frac{y - x'\hat \beta_j}{\exp (x_{2j}'\hat \gamma_j/2)} \right\},\label{sca:dist}
\end{equation}
where $\hat \beta_j$ is the least square estimator (\ref{eq:ols}), $X_{2j} \subseteq X_{j}$ with $\dim X_{2j} = d_{x_2},$ and
$$
\hat \gamma_j = \arg \min_{g \in \mathbb{R}^{d_{x_2}}} \sum_{i=1}^{n_j}  (\log (Y_{ji} - X_{ji}'\hat \beta_j)^2 - X_{2ji}'g)^2.
$$
The option \emph{scale\_variable} specifies the covariates $X_{2j}$ that affect the scale of the conditional distribution. The option \emph{counterfactual\_scale\_variable} selects the counterfactual scale variables when the counterfactual population is specified using \emph{counterfactual\_var}. By default, \emph{R} would use all the covariates as \emph{scale\_variable} and  \emph{counterfactual\_scale\_variable = counterfactual\_var}.  The estimator (\ref{sca:dist}) is based on a restrictive location scale shift model that imposes that the covariates $X$ only affect the location and scale of the outcome $Y$.

\item \emph{method = "cqr"} implements the censored quantile regression estimator of the conditional distribution, which is the same as (\ref{qr:dist}) with $\hat \beta(u)$ replaced by the \citep{Chernozhukov2002} censored quantile regression estimator. The options \emph{trimming} and \emph{nreg} apply to this method with the same functionality as for the \emph{qr} method. Moreover,  a variable containing a censoring indicator $C_j$ must be specified with \emph{censoring}. The censored quantile regression estimator has three-steps by default. The number of steps can be increased by the option \emph{nsteps}. In the first step, the censoring probabilities are estimated by a logit regression of the censoring indicator $C_j$ on all the covariates $X_j$. Then, for each quantile index $u$, the observations with sufficiently low censoring probabilities relative to $u$ are selected. We allow for misspecification of the logit by excluding the observations that could theoretically be used but have censoring probabilities in the highest \emph{firstc} quantiles, with a default of $0.1$, i.e. $10\%$ of the observations. In the second step, standard linear quantile regressions are estimated on the samples defined in step one. Using the estimated quantile regressions, we define a new sample of observations that can be used. This sample consists of all observations for which the estimated conditional quantile is above the censoring point. Again, we throw away observations in the lowest \emph{secondc} quantiles of the distribution of the residuals, with a default of $0.05$, i.e. $5\%$ of the observations. Step three consists in a new linear quantile regression using the sample defined in step two. Step three is repeated if \emph{nsteps} is above 3. 
This method should be used only with censored dependent variables.

\item \emph{method = "cox"} implements the duration regression estimator of the conditional distribution function
\begin{equation}
\hat F_{Y_{j}|X_{j}}(y|x) = 1 - \exp(-\exp(\hat t(y) - x'\hat \beta)),\label{cox:dist}
\end{equation}
where $\hat \beta$ is the Cox estimator of the regression coefficients and $\hat t(y)$ is the Cox estimator of the baseline integrated hazard function \citep{Cox1972}. The Cox estimator calls the R package \CRANpkg{survival} \citep{survival-package}. The estimator (\ref{cox:dist}) is based on a restrictive transformation location shift model that imposes that the covariates $X$ only affect the location of a monotone transformation of the outcome $t(Y)$, i.e.
$$
t(Y_j) = X_j'\beta_j + V_j,
$$
where $V_j$ has an extreme value distribution and is independent of $X_j$. This method should be used only with nonnegative dependent variables.

\item \emph{method = "logit"} implements the distribution regression estimator of the conditional distribution with logistic link function
\begin{equation}
\hat F_{Y_{j}|X_{j}}(y|x) = \Lambda(x'\hat \beta(y)),\label{dr:dist}
\end{equation}
where $\Lambda$ is the standard logistic distribution function, and $\hat \beta(y)$ is the distribution regression estimator
\begin{equation}
\hat \beta(y) = \arg \max_{b \in \mathbb{R}^{d_x}} \sum_{i=1}^{n_j} \left[1\{Y_{ji} \leq y\} \log \Lambda(X_{ij}'b) + 1\{Y_{ij} > y\} \log \Lambda(-X_{ji}'b) \right]. \label{eq:dr}
\end{equation}
The estimator (\ref{dr:dist}) is based on a flexible model where each covariate can have a heterogenous effect at different parts of the distribution. This method can be used with continuous dependent variables and censored dependent variables with fixed censoring point.

\item \emph{method = "probit"} implements the distribution regression estimator of the conditional distribution with normal link function, i.e. where $\Lambda$ is the standard normal distribution function in (\ref{dr:dist}) and (\ref{eq:dr}).

\item \emph{method = "lpm"} implements the linear probability model estimator of the conditional distribution 
\[
\hat F_{Y_{j}|X_{j}}(y|x) = x'\hat \beta(y),
\]
where $\hat \beta(y)$ is the least squares estimator
\[
\hat \beta(y) = \arg \min_{b \in \mathbb{R}^{d_x}} \sum_{i=1}^{n_j} (1\{Y_{ji} \leq y\} -X_{ij}'b )^2. 
\]
This method might produce estimates of the conditional distribution outside the interval $[0,1]$.

\end{enumerate}

For the methods (2)--(8), the option \emph{nreg} sets the number of values of $y$ to evaluate the estimator of the conditional distribution function. These values are uniformly distributed among the observed values of $Y_j$. If \emph{nreg} is greater than the number of observed values of $Y_j$, then all the observed values are used.

\subsection{Inference}

The command \emph{counterfactual} reports pointwise and uniform confidence intervals for the QEs over a prespecified set of quantile indexes. The construction of the intervals rely on  functional central limit theorems and bootstrap functional central limit theorems for the empirical QEs derived in \citep{ChernozhukovFernandez-ValMelly2013}. In particular, the pointwise intervals are based on the normal distribution, whereas the uniform intervals are based on two resampling schemes: empirical and weighted bootstrap. Thus, the $(1-\alpha)$ confidence interval for $\Delta(\tau)$ on $\mathcal{T}$ has the form
$$
\{ \hat \Delta(\tau) \pm c_{1-\alpha} \hat \Sigma(\tau) : \tau \in \mathcal{T}\},
$$
where $\hat \Sigma(\tau)$ is the standard error of $\hat \Delta(\tau)$ and $c_{1-\alpha}$ is a critical value. There are two options to obtain $\hat \Sigma(\tau)$. The default option \emph{robust=FALSE} computes the bootstrap standard deviation of $\hat \Delta(\tau)$; whereas the option \emph{robust=TRUE} computes the bootstrap interquartile range rescaled with the normal distribution, $\hat \Sigma(\tau) = (q_{0.75}(\tau) - q_{0.25}(\tau))/(z_{0.75} - z_{0.25})$ where $q_p(\tau)$ is the $p$th quantile of the bootstrap draws of $\hat \Delta(\tau)$ and $z_p$ is the $p$th quantile of the standard normal. The pointwise critical value is $c_{1-\alpha} = z_{1-\alpha}$, and the uniform critical 
value is $c_{1-\alpha} = \hat t_{1-\alpha},$ where $\hat t_{1-\alpha}$ is a bootstrap estimator of the $(1-\alpha)$th quantile of the Kolmogorov-Smirnov maximal t-statistic 
$$
t = \sup_{\tau \in \mathcal{T}} |\hat \Delta(\tau) - \Delta(\tau) | / \hat \Sigma(\tau).
$$

\medskip

In addition to the intervals, \emph{counterfactual} reports the p-values for several functional tests based on two test-statistic: Kolmogorov-Smirnov and the Cramer-von-Misses-Smirnov. The null-hypotheses considered are
\begin{enumerate}
\item Correct parametric specification of the model for the conditional distribution.  This test compares the empirical distribution of the outcome $Y_j$ with the estimate of the counterfactual distribution in the reference population 
\[
\hat F_{Y\left\langle j|j\right\rangle }(y):=\int_{\mathcal{X}_{j}}\hat F_{Y_{j}|X_{j}}(y|x)d\hat F_{X_{j}}(x).
\]
The power of this specification test might be low  because it only uses the implications of the conditional distribution on the counterfactual distribution. For example, the test is not informative for the linear probability and logit models where the counterfactual distribution in the reference population is identical to the empirical distribution by construction. If \emph{group} is specified and \emph{treatment=TRUE} is selected, then the test is performed in the population defined by \emph{group=1}. If in addition the option \emph{decomposition=TRUE} is selected, then the test is performed in the populations defined by \emph{group=0} and \emph{group=1}, and in the combined population including both  \emph{group=0} and \emph{group=1}.

\item Zero QE at all the quantile indexes of interest: $\Delta(\tau) = 0$ for all  $\tau \in \mathcal{T}$. This is stronger than a zero average effect. Other null hypotheses of constant quantile effect (but at a different level than 0) can be added with the option \emph{cons\_test}. 

\item Constant QE at all quantile indexes of interest: $\Delta(\tau) = \Delta(0.5)$ for all  $\tau \in \mathcal{T}$. 

\item First-order stochastic dominance: $\Delta(\tau) \geq 0$ for all  $\tau \in \mathcal{T}$.

\item Negative first-order stochastic dominance: $\Delta(\tau) \leq 0$ for all  $\tau \in \mathcal{T}$.

\end{enumerate}

\medskip

The options of  \emph{counterfactual} related to inference are:

\begin{enumerate}

\item \emph{noboot = TRUE} suppresses the bootstrap. The bootstrap can be very demanding in terms of computation time. Therefore, it is recommended to switch it off for the exploratory analysis of the data.

\item \emph{weightedboot = TRUE} selects weighted bootstrap with standard exponential weights. The default \emph{weightedboot = FALSE} selects empirical bootstrap with multinomial weights. We recommend weighted bootstrap when the covariates include categorical variables with small cell sizes to avoid singular designs in the bootstrap draws. 

\item \emph{reps} specifies the number of bootstrap replications. This number will matter only if the bootstrap has not be suppressed. The default is 100.

\item \emph{alpha} specifies the significance level of the tests and confidence intervals. Note that the confidence level of the confidence interval is 1 - \emph{alpha}. Thus, the default value of $0.05$ produces $95\%$ confidence intervals.

\item \emph{first} and \emph{last} select the subset of quantile indexes of interest for inference.  The tails of the distribution should not be used because standard asymptotic does not apply to these parts.  The needed amount of tail trimming depends on the sample size and on the distribution of the dependent variable.  \emph{first} sets the lowest quantile index used and \emph{last} sets the highest quantile index used.  The default values are 0.1 and 0.9 so that $\mathcal{T} = [0.1, 0.9]$.

\item \emph{cons\_test} add tests of the null hypothesis that $\Delta(\tau)  = $ \emph{const\_test} for all $\tau$ between \emph{first} and \emph{last}.  The null hypothesis that $\Delta(\tau)  = 0$ for all  $\tau$ between \emph{first} and \emph{last} is tested by default.  The null hypothesis that the quantile effects are constant is also tested by default.

\end{enumerate}

\subsection{Parallel Computing}
The command \emph{counterfactual} provides functionality for parallel computing, which is specially useful to speed up the execution of the bootstrap. There are two options related to parallel computing:

\begin{enumerate}
\item \emph{setcore} specifies whether multiple cores should be used. The default value \emph{setcore = FALSE} turns off the parallel computing.
\item \emph{ncore} selects the number of cores to use for parallel computing. The information of this option is only used when parallel computing is switched on with \emph{setcore = TRUE}.
\end{enumerate}

\section{Empirical Examples}
We consider two empirical examples to illustrate the functionality of the command \emph{counterfactual}. The first example is an estimation of Engel curves that includes a counterfactual analysis where the counterfactual population is an artificial transformation of a reference population. The second example is wage decomposition with respect to union status where the reference and counterfactual populations correspond to two different  groups. 

\subsection{Engel Curves}\label{Sec:example2}
We use the classical Engel 1857 dataset  to estimate the relationship between food expenditure (\textit{foodexp}) and annual household income (\textit{income}), and then report the estimates of the QE of a change in the distribution of the annual household income that might be induced for example by a variation in income taxation.\footnote{This is the same data set as in the quantile regression package \CRANpkg{quantreg}, see
\citep{quantreg-package}.}
We estimate the conditional distribution with the quantile regression method, i.e., \emph{method ="qr"}.

\medskip

First, we generate the variable \textit{counterfactual\_income} with the counterfactual values of income and plot the reference and counterfactual income distributions. The counterfactual distribution corresponds to a mean preserving spread of the distribution in the reference population that reduces standard deviation by $25\%$. 

\begin{Schunk}
\begin{Sinput}
> library(quantreg)
> data(engel)
> attach(engel)
> counter_income <- mean(income)+0.75*(income-mean(income))
> cdfx <- c(1:length(income))/length(income)
> plot(c(0,4000),range(c(0,1)), xlim =c(0, 4000), type="n", xlab = "Income", 
+      ylab="Probability")
> lines(sort(income), cdfx)
> lines(sort(counter_income), cdfx, lwd = 2, col = 'grey70')
> legend(1500, .2, c("Original", "Counterfactual"), lwd=c(1,2),bty="n", 
+        col=c(1,'grey70'))
\end{Sinput}
\end{Schunk}

\begin{figure} [htbp]
\centering
\includegraphics[width=0.7\textwidth]{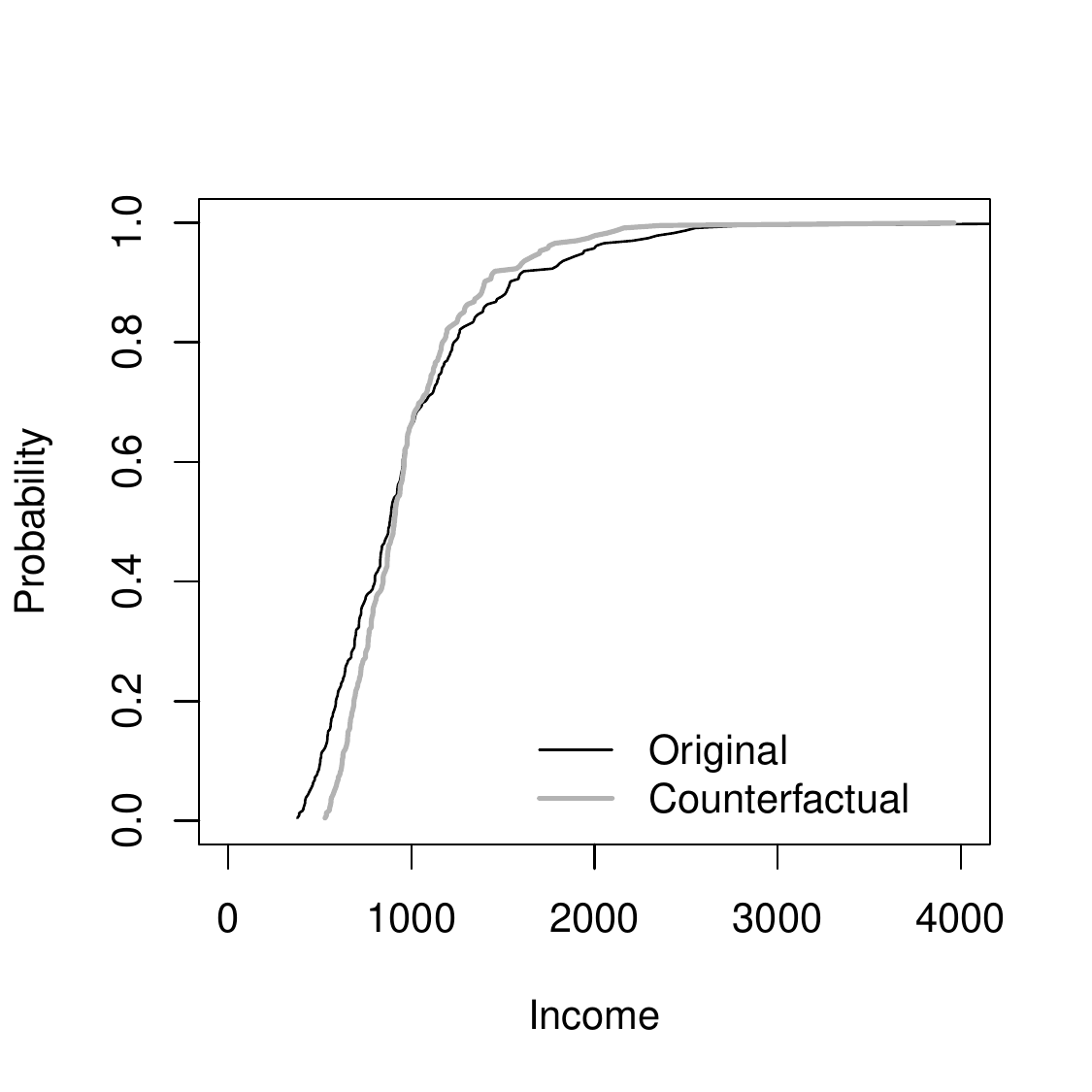}
\caption{Observed and counterfactual distributions of income}
\end{figure}

\medskip

To estimate the QEs of this counterfactual change we turn on the option \textit{transformation} of \textit{counterfactual} by setting \textit{transformation = TRUE}, and specify that the counterfactual values of the covariate \textit{income} are in the generated variable \textit{counter\_income} by setting \textit{counterfactual\_var = counter\_income}. This yields:

\begin{Schunk}
\begin{Sinput}
> qrres <- counterfactual(foodexp~income, counterfactual_var
+         = counter_income, transformation = TRUE)
\end{Sinput}

\begin{Soutput}
Conditional Model:                      linear quantile regression
Number of regressions estimated:         100 

The variance has been estimated by bootstraping the results 100 times.

No. of obs. in the reference group:      235 
No. of obs. in the counterfactual group: 235 


                   Quantile Effects -- Composition                     
---------------------------------------------------------------------- 
                     Pointwise      Pointwise           Functional     
 Quantile     Est.    Std.Err   95% Conf.Interval   95% Conf.Interval  
     0.1    68.049     4.214    59.789    76.308    55.939    80.159
     0.2    57.897     4.332    49.406    66.388    45.448    70.346
     0.3    43.851     5.246    33.568    54.133    28.774    58.927
     0.4    29.248     5.091    19.270    39.227    14.618    43.878
     0.5    16.716     4.602     7.696    25.735     3.491    29.940
     0.6     5.744     4.308    -2.698    14.187    -6.634    18.123
     0.7    -8.866     7.132   -22.845     5.113   -29.361    11.630
     0.8   -40.099     8.191   -56.153   -24.045   -63.637   -16.561
     0.9    -88.56     13.83   -115.67    -61.44   -128.31    -48.80


      Bootstrap inference on the counterfactual quantile process       
---------------------------------------------------------------------- 
						    P-values	
					        ====================== 
NULL-Hypthoesis                        	 KS-statistic CMS-statistic 
====================================================================== 
Correct specification of the parametric model          0          0 
No effect: QE(tau)=0 for all taus                      0          0 
Constant effect: QE(tau)=QE(0.5) for all taus          0          0 
Stochastic dominance: QE(tau)>0 for all taus           0          0 
Stochastic dominance: QE(tau)<0 for all taus           0          0 
\end{Soutput}
\end{Schunk}

We reject the simultaneous hypotheses of zero, constant, positive and negative effect of the income redistribution at all the deciles. The QR model for the conditional distribution cannot be rejected at conventional significance levels. 

\begin{figure}[htbp]
\centering
\includegraphics[width=0.7\textwidth]{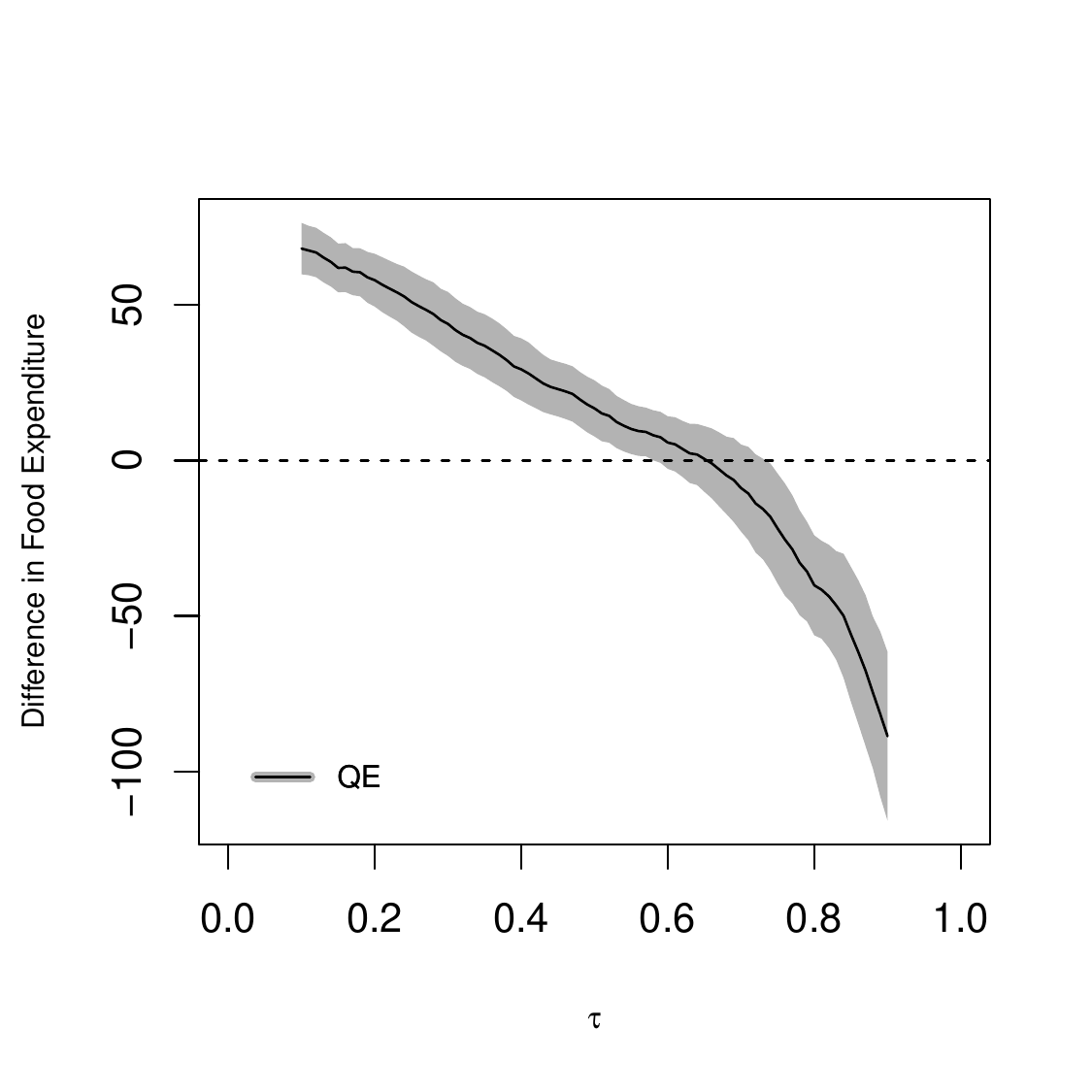}
\caption{Quantile effects of income redistribution on food consumption}\label{fig:engel}
\end{figure}

\medskip
Finally, we reestimate the QE function on the larger set of quantiles $\{0.01, 0.02, \ldots , 0.99 \}$, and plot a uniform confidence band over the subset $\{0.10, 0.11, \ldots , 0.90 \}$ constructed by empirical bootstrap with 100 replications. In Figure \ref{fig:engel} we can visually reject the functional hypotheses of zero, constant, positive and negative effect at the percentiles considered.  We use the option \emph{printdeco = FALSE} to suppress the display of the table of results.

\begin{Schunk}
\begin{Sinput}
> taus  <- c(1:99)/100
> first <- sum(as.double(taus <= .10))
> last  <- sum(as.double(taus <= .90))
> rang  <- c(first:last)
> rqres <- counterfactual(foodexp~income, counterfactual_var=counter_income, 
+                         nreg=100, quantiles=taus, transformation = TRUE, 
+                         printdeco = FALSE, sepcore = TRUE,ncore=2)
\end{Sinput}

\begin{Soutput}
cores used= 2 
\end{Soutput}

\begin{Sinput}
> duqf   <- (rqres$resCE)[,1]
> l.duqf <- (rqres$resCE)[,3]
> u.duqf <- (rqres$resCE)[,4]
\end{Sinput}
\end{Schunk}

\begin{Schunk}
\begin{Sinput}
> plot(c(0,1), range(c(min(l.duqf[rang]), max(u.duqf[rang]))), xlim = c(0,1), 
+      type = "n", xlab = expression(tau), ylab = "Difference in Food Expenditure", 
+      cex.lab=0.75)
> polygon(c(taus[rang], rev(taus[rang])), c(u.duqf[rang], rev(l.duqf[rang])), 
+         density = -100, border = F, col = "grey70", lty = 1, lwd = 1)
> lines(taus[rang], duqf[rang]) 
> abline(h = 0, lty = 2) 
> legend(0, -90, "QE", cex = 0.75, lwd = 4, bty = "n", col = "grey70")
> legend(0, -90, "QE", cex = 0.75, lty = 1, bty = "n", lwd = 1)
\end{Sinput}
\end{Schunk}

\subsection{Union Premium}\label{sec:examples}
We use an extract of the U.S. National Longitudinal Survey of Young Women (NLSW) for employed women in 1988 to estimate a wage decomposition with respect to union status.\footnote{This dataset is available from the Stata's sample datasets at \textit{http://www.stata-press.com/data/r9/nlsw88.dta}.} The outcome variable $Y$ is the log hourly wage (\textit{lwage}), the covariates $X$ include job tenure in years (\textit{tenure}), years of schooling (\textit{grade}), and total experience (\textit{ttl\_exp}), and the union indicator \textit{union} defines the reference and counterfactual populations. We estimate the conditional distributions by distribution regression with logistic link and duration regression, i.e., \emph{method ="logit"} and \emph{method ="cox"}. We use weighted bootstrap for the construction of uniform confidence intervals and hypothesis tests and run  parallel computing with 2 nodes.

\medskip

We start by estimating the wage decomposition by logistic distribution regression, where the counterfactual population is specified with \emph{group=union} with the options \emph{treatment=TRUE} and \emph{decomposition=TRUE} to estimate the composition, structure and total effects. The structure effect in this case correspond to the treatment effect of union on the treated or union premium. The tables show that the union workers earn higher wages than the nonumion workers throuoghout the distribution although the union wage gap is decreasing in the quantile index. This gap can be mostly explained by differences in tenure, education and experience between union and nonunion workers in the upper tail of the distribution and by the union premium in the rest of the distribution.

\begin{Schunk}
\begin{Sinput}
> data(nlsw88)
> attach(nlsw88)
> lwage    <- log(wage)
> logitres <- counterfactual(lwage~tenure+ttl_exp+grade, 
+                           group = union, treatment=TRUE,  
+                           decomposition=TRUE, method = "logit", 
+                           weightedboot = TRUE, sepcore = TRUE, ncore=2)
\end{Sinput}
\begin{Soutput}
cores used= 2 
                                  
Conditional Model:                      logit
Number of regressions estimated:         96 

The variance has been estimated by bootstraping the results 100 times.

No. of obs. in the reference group:      1407 
No. of obs. in the counterfactual group: 459 


                    Quantile Effects -- Structure                      
---------------------------------------------------------------------- 
                     Pointwise      Pointwise           Functional     
 Quantile     Est.    Std.Err   95% Conf.Interval   95% Conf.Interval  
     0.1   0.24047   0.06005   0.12278   0.35817   0.07757   0.40338
     0.2   0.21903   0.05630   0.10869   0.32937   0.06631   0.37176
     0.3   0.23437   0.04628   0.14366   0.32508   0.10881   0.35992
     0.4   0.18524   0.04252   0.10190   0.26857   0.06989   0.30059
     0.5   0.16041   0.04404   0.07410   0.24671   0.04095   0.27987
     0.6   0.13897   0.04618   0.04845   0.22949   0.01369   0.26425
     0.7   0.05701   0.04407  -0.02937   0.14339  -0.06255   0.17657
     0.8   0.01945   0.04179  -0.06245   0.10135  -0.09391   0.13281
     0.9  0.006434  0.078547 -0.147514  0.160382 -0.206650  0.219518


      Bootstrap inference on the counterfactual quantile process       
---------------------------------------------------------------------- 
						    P-values	
					        ====================== 
NULL-Hypthoesis                        	 KS-statistic CMS-statistic 
====================================================================== 
Correct specification of the parametric model          1          1 
No effect: QE(tau)=0 for all taus                      0          0 
Constant effect: QE(tau)=QE(0.5) for all taus          0       0.02 
Stochastic dominance: QE(tau)>0 for all taus        0.95       0.95 
Stochastic dominance: QE(tau)<0 for all taus           0          0 


                   Quantile Effects -- Composition                     
---------------------------------------------------------------------- 
                     Pointwise      Pointwise           Functional     
 Quantile     Est.    Std.Err   95% Conf.Interval   95% Conf.Interval  
     0.1   0.06062   0.04131  -0.02035   0.14160  -0.05752   0.17877
     0.2   0.04879   0.03717  -0.02406   0.12164  -0.05750   0.15508
     0.3   0.05313   0.03721  -0.01981   0.12607  -0.05329   0.15956
     0.4   0.09245   0.03772   0.01851   0.16638  -0.01543   0.20033
     0.5   0.08952   0.03945   0.01220   0.16683  -0.02329   0.20233
     0.6   0.12259   0.03862   0.04690   0.19828   0.01215   0.23303
     0.7   0.12975   0.03781   0.05564   0.20385   0.02162   0.23787
     0.8  0.090722  0.030184  0.031563  0.149881  0.004404  0.177041
     0.9   0.05503   0.05744  -0.05755   0.16760  -0.10924   0.21929


      Bootstrap inference on the counterfactual quantile process       
---------------------------------------------------------------------- 
						    P-values	
					        ====================== 
NULL-Hypthoesis                        	 KS-statistic CMS-statistic 
====================================================================== 
Correct specification of the parametric model          1          1 
No effect: QE(tau)=0 for all taus                   0.01       0.01 
Constant effect: QE(tau)=QE(0.5) for all taus       0.77       0.58 
Stochastic dominance: QE(tau)>0 for all taus        0.81       0.81 
Stochastic dominance: QE(tau)<0 for all taus        0.01       0.01 


                      Quantile Effects -- Total                        
---------------------------------------------------------------------- 
                     Pointwise      Pointwise           Functional     
 Quantile     Est.    Std.Err   95% Conf.Interval   95% Conf.Interval  
     0.1   0.30110   0.05703   0.18933   0.41287   0.13655   0.46565
     0.2   0.26782   0.06383   0.14271   0.39293   0.08363   0.45202
     0.3   0.28750   0.05459   0.18051   0.39449   0.12998   0.44502
     0.4   0.27768   0.05211   0.17556   0.37981   0.12733   0.42804
     0.5   0.24992   0.05111   0.14975   0.35010   0.10244   0.39741
     0.6   0.26156   0.04999   0.16358   0.35953   0.11731   0.40580
     0.7   0.18675   0.04529   0.09800   0.27551   0.05608   0.31743
     0.8   0.11017   0.04624   0.01954   0.20081  -0.02327   0.24361
     0.9   0.06146   0.06287  -0.06176   0.18468  -0.11996   0.24287


      Bootstrap inference on the counterfactual quantile process       
---------------------------------------------------------------------- 
						    P-values	
					        ====================== 
NULL-Hypthoesis                        	 KS-statistic CMS-statistic 
====================================================================== 
Correct specification of the parametric model          1          1 
No effect: QE(tau)=0 for all taus                      0          0 
Constant effect: QE(tau)=QE(0.5) for all taus       0.06       0.13 
Stochastic dominance: QE(tau)>0 for all taus        0.93       0.93 
Stochastic dominance: QE(tau)<0 for all taus           0          0 
\end{Soutput}
\end{Schunk}
\medskip

\begin{figure}[htbp]
\includegraphics{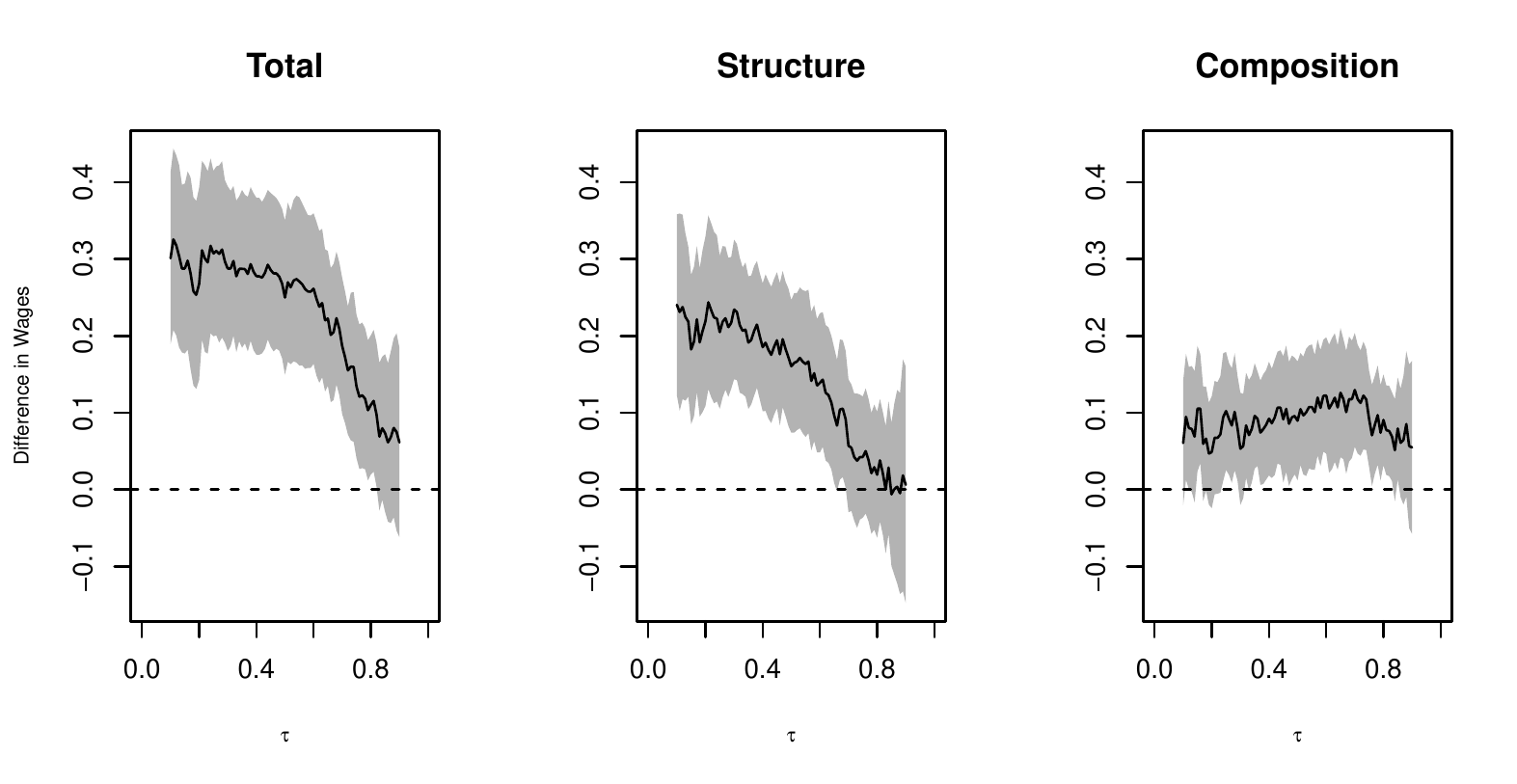}
\caption{Wage decomposition with respect to union: logit regression estimates}\label{fig:union}
\end{figure}

Next, we reestimate the QE function on the larger set of quantiles $\{0.01, 0.02, ..., 0.99 \}$, and plot a uniform confidence band over the subset $\{0.10, 0.11, ..., 0.90 \}$) constructed by weighted bootstrap with 100 replications. Figure \ref{fig:union} shows  that the structure effect is heterogeneous across the quantile indexes and explains most of the union wage gap below the third quartile. The composition effect is constant across quantile indexes and explains most of the wage gap above the third quartile. 

\begin{Schunk}
\begin{Sinput}
> taus  <- c(1:99)/100
> first <- sum(as.double(taus <= .10))
> last  <- sum(as.double(taus <= .90))
> rang  <- c(first:last) 
> logitres <- counterfactual(lwage~tenure+ttl_exp+grade, 
+           group = union, treatment=TRUE, quantiles=taus, 
+           method="logit", nreg=100, weightedboot = TRUE, 
+           printdeco=FALSE, decomposition = TRUE, 
+           sepcore = TRUE,ncore=2)
\end{Sinput}
\begin{Soutput}
cores used= 2 
\end{Soutput}
\begin{Sinput}
> duqf_SE   <- (logitres$resSE)[,1]
> l.duqf_SE <- (logitres$resSE)[,3]
> u.duqf_SE <- (logitres$resSE)[,4]
> duqf_CE   <- (logitres$resCE)[,1]
> l.duqf_CE <- (logitres$resCE)[,3]
> u.duqf_CE <- (logitres$resCE)[,4]
> duqf_TE   <- (logitres$resTE)[,1]
> l.duqf_TE <- (logitres$resTE)[,3]
> u.duqf_TE <- (logitres$resTE)[,4]
> range_x <- min(c(min(l.duqf_SE[rang]), min(l.duqf_CE[rang]), 
+                 min(l.duqf_TE[rang])))
> range_y <- max(c(max(u.duqf_SE[rang]), max(u.duqf_CE[rang]),
+                 max(u.duqf_TE[rang])))
\end{Sinput}
\end{Schunk}

\begin{Schunk}
\begin{Sinput}
> par(mfrow=c(1,3))
> plot(c(0,1), range(c(range_x, range_y)), xlim = c(0,1), type = "n", 
+      xlab = expression(tau), ylab = "Difference in Wages", cex.lab=0.75,  
+      main = "Total")
> polygon(c(taus[rang],rev(taus[rang])), 
+         c(u.duqf_TE[rang], rev(l.duqf_TE[rang])), density = -100, border = F, 
+         col = "grey70", lty = 1, lwd = 1)
> lines(taus[rang], duqf_TE[rang]) 
> abline(h = 0, lty = 2)
> plot(c(0,1), range(c(range_x, range_y)), xlim = c(0,1), type = "n", 
+      xlab = expression(tau), ylab = "", cex.lab=0.75, main = "Structure")
> polygon(c(taus[rang],rev(taus[rang])), 
+         c(u.duqf_SE[rang], rev(l.duqf_SE[rang])), density = -100, border = F, 
+         col = "grey70", lty = 1, lwd = 1)
> lines(taus[rang], duqf_SE[rang]) 
> abline(h = 0, lty = 2)
> plot(c(0,1), range(c(range_x, range_y)), xlim = c(0,1), type = "n", 
+      xlab = expression(tau), ylab = "", cex.lab=0.75, main = "Composition")
> polygon(c(taus[rang],rev(taus[rang])), 
+         c(u.duqf_CE[rang], rev(l.duqf_CE[rang])), density = -100, border = F, 
+         col = "grey70", lty = 1, lwd = 1)
> lines(taus[rang], duqf_CE[rang]) 
> abline(h = 0, lty = 2)
\end{Sinput}
\end{Schunk}

\begin{figure}[htbp]
\includegraphics{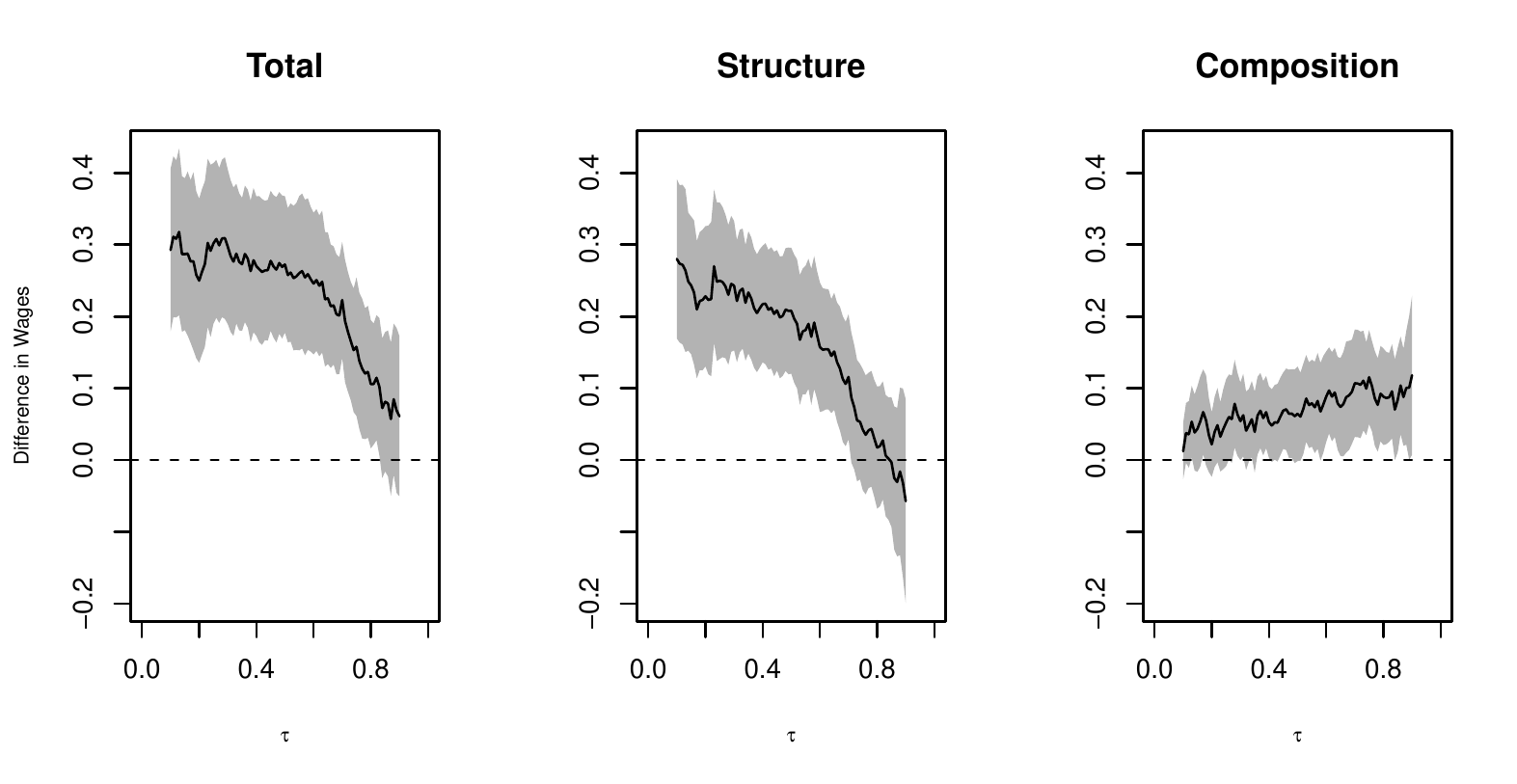}
\caption{Wage decomposition with respect to union: duration regression estimates}\label{fig:union2}
\end{figure}

\medskip
Finally, we repeat the point and interval estimation using the duration regression method with the option \emph{method = "cox"}. Despite of relying on a more restrictive model for the conditional distribution, the duration regression estimates in Figure \ref{fig:union2}  are similar to the logit regression estimates in Figure \ref{fig:union}.

\begin{Schunk}
\begin{Sinput}
> coxres <- counterfactual(lwage~tenure+ttl_exp+grade, 
+           group = union, treatment=TRUE, quantiles=taus,  
+           method="cox", nreg=100, weightedboot = TRUE,  
+           printdeco = FALSE, decomposition = TRUE, sepcore = TRUE,ncore=2)
\end{Sinput}
\begin{Soutput}
cores used= 2 
\end{Soutput}
\begin{Sinput}
> duqf_SE   <- (coxres$resSE)[,1]
> l.duqf_SE <- (coxres$resSE)[,3]
> u.duqf_SE <- (coxres$resSE)[,4]
> duqf_CE   <- (coxres$resCE)[,1]
> l.duqf_CE <- (coxres$resCE)[,3]
> u.duqf_CE <- (coxres$resCE)[,4]
> duqf_TE   <- (coxres$resTE)[,1]
> l.duqf_TE <- (coxres$resTE)[,3]
> u.duqf_TE <- (coxres$resTE)[,4]
> range_x = min(c(min(l.duqf_SE[rang]), min(l.duqf_CE[rang]), 
+                 min(l.duqf_TE[rang])))
> range_y = max(c(max(u.duqf_SE[rang]), max(u.duqf_CE[rang]), 
+                 max(u.duqf_TE[rang])))
\end{Sinput}
\end{Schunk}

\begin{Schunk}
\begin{Sinput}
> par(mfrow=c(1,3))
> plot(c(0,1), range(c(range_x, range_y)), xlim = c(0,1), type = "n", 
+      xlab = expression(tau), ylab = "Difference in Wages", cex.lab=0.75, 
+      main = "Total")
> polygon(c(taus[rang],rev(taus[rang])), 
+         c(u.duqf_TE[rang], rev(l.duqf_TE[rang])), density = -100, border = F, 
+         col = "grey70", lty = 1, lwd = 1)
> lines(taus[rang], duqf_TE[rang]) 
> abline(h = 0, lty = 2)
> plot(c(0,1), range(c(range_x, range_y)), xlim = c(0,1), type = "n", 
+      xlab = expression(tau), ylab = " ", cex.lab=0.75, main = "Structure")
> polygon(c(taus[rang],rev(taus[rang])), 
+         c(u.duqf_SE[rang], rev(l.duqf_SE[rang])), density = -100, border = F, 
+         col = "grey70", lty = 1, lwd = 1)
> lines(taus[rang], duqf_SE[rang]) 
> abline(h = 0, lty = 2)
> plot(c(0,1), range(c(range_x, range_y)), xlim = c(0,1), type = "n", 
+      xlab = expression(tau), ylab = "", cex.lab=0.75, main = "Composition")
> polygon(c(taus[rang],rev(taus[rang])), 
+         c(u.duqf_CE[rang], rev(l.duqf_CE[rang])), density = -100, border = F, 
+         col = "grey70", lty = 1, lwd = 1)
> lines(taus[rang], duqf_CE[rang]) 
> abline(h = 0, lty = 2)
\end{Sinput}
\end{Schunk}



\pagebreak
\bibliography{chen-chernozhukov-fernandezval-melly}

\address{Mingli Chen\\
  Department of Economics\\
  University of Warwick\\
  UK\\}
\email{m.chen.3@warwick.ac.uk}

\address{Victor Chernozhukov\\
 Department of Economics\\
 Massachusetts Institute of Technology\\
  USA\\}
\email{vchern@mit.edu}

\address{Iván Fernández-Val\\
  Department of Economics\\
  Boston University\\
  USA\\}
\email{ivanf@bu.edu}

\address{Blaise Melly\\
  Department of Economics \\
  University of Bern \\
  Switzerland\\}
\email{mellyblaise@gmail.com}
\end{article}

\end{document}